\documentclass{article}
\pdfoutput=1
\usepackage{graphicx}
\usepackage{booktabs}
\usepackage{authblk} 
\usepackage{amsmath}
\usepackage{amssymb}
\usepackage{xcolor}
\usepackage{multicol}
\usepackage{geometry}
\usepackage{{float}}
\usepackage{xr} 
\makeatletter
\newcommand*{\addFileDependency}[1]{
\typeout{(#1)}
\IfFileExists{#1}{}{\typeout{No file #1.}}
}\makeatother

\newcommand*{\myexternaldocument}[1]{%
\externaldocument{#1}%
\addFileDependency{#1.tex}%
\addFileDependency{#1.aux}%
}
\myexternaldocument{supplementary_material}
\usepackage[hidelinks]{hyperref}
\usepackage[capitalise]{cleveref}

\usepackage[numbers,sort&compress]{natbib}
\usepackage[acronym]{glossaries}
\makeglossaries
\newacronym{SPH}{SPH}{smoothed particle hydrodynamics}
\newacronym{XFEM}{XFEM}{extended finite element method}
\newacronym{CZM}{CZM}{cohesive-zone model}
\newacronym{LCP}{LCP}{linear complementarity problem}
\newacronym{MLCP}{MLCP}{mixed linear complementarity problem}
\newacronym{SDOF}{SDOF}{single degree-of-freedom}
\newacronym{DOF}{DOF}{degree-of-freedom}
\newacronym{TSL}{TSL}{traction-separation law}
\newacronym{CD}{CD}{central-difference}
\newacronym{CFL}{CFL}{Courant-Friedrichs-Lewy}
\newacronym{FEM}{FEM}{finite-element method}

\title{Stability of Extrinsic Cohesive-Zone Model with Penalty-Based Contact in Explicit Dynamic Fragmentation Simulations}

\author[1]{Thibault Ghesquière-Diérickx}
\author[1]{Jean-François Molinari}
\author[1]{Guillaume Anciaux\thanks{Corresponding author: \href{mailto:guillaume.anciaux@epfl.ch}{guillaume.anciaux@epfl.ch}}}
\affil[1]{Institute of Civil Engineering, Institute of Materials Science and Engineering, École Polytechnique Fédérale de Lausanne (EPFL), Lausanne, Switzerland}

\date{}

\begin{document}

\maketitle

\begin{abstract}

    \noindent
    Dynamic fragmentation simulations are essential for predicting material response at high strain rates, yet explicit dynamic simulations that combine an extrinsic \acrfull{CZM} with penalty-based contact often exhibit severe instabilities. In a two-dimensional benchmark, we observe exponential energy growth and resulting artificial fragmentation under standard contact penalty settings and time step choices, which motivates a systematic analysis of instability sources. Three mechanisms are isolated and quantified: (i) diverging initial cohesive stiffness, which constrains the stable time step; (ii) discontinuous stiffness jumps at the cohesive–contact interface; and (iii) discontinuity introduced by cohesive softening. Analytical error estimates, phase-space diagnostics, and energy growth metrics reveal that repeated cohesive–contact switching can accumulate small per-step energy errors into long-term energy drift. Within the explored parameter space, maintaining stability requires time steps well below the usual limit. To mitigate these energy artifacts, we assess an adaptive penalty strategy that ties the contact stiffness to the evolving cohesive stiffness. This modification eliminates the discontinuity and restores energy conservation, but it allows larger interpenetration, making it suitable as a diagnostic rather than a definitive remedy. Overall, our study identifies the root causes of unphysical energy drift and demonstrates that penalty-based contact is not a viable approach for long-term, energy-consistent fragmentation simulations with physically meaningful fragment statistics.

    \vspace{0.5cm}
    \noindent\textbf{Keywords:} cohesive‐zone model; penalty methods; explicit dynamics; dynamic fragmentation; numerical stability; contact mechanics; energy conservation; nonsmooth mechanics

\end{abstract}



\section{Introduction}

Hypervelocity collisions between orbital assets (satellites, mission hardware, legacy debris) generate thousands of fragments in milliseconds, posing a critical threat to spacecraft operations and long-term orbital sustainability \citep{kessler1978collision,liou2006risks}. These high-energy events not only elevate the risk of further collisions (the so-called debris cascade or "Kessler" effect) but also generate fragment size and velocity distributions that feed debris-environment models and collision-avoidance algorithms. Accurate quantification of these distributions is therefore crucial for risk assessment and mitigation planning \citep{klinkrad2006space}.

\medskip\noindent
Experimental campaigns such as SOCIT \citep{krisko2008socit4} and DebriSat \citep{liou2013debrisat, cowardin2023updates}, along with observational data, have highlighted the complex physics of these events, where materials under extreme strain rates behave more like fluids than solids. At the material scale, theoretical work provides a mechanistic understanding of how pre-existing defects trigger crack nucleation, growth, branching, and coalescence into fragments that may subsequently collide \citep{grady1985mechanisms, grady2007fragmentation, ramesh2015review, hild2015characteristic}. This motivates numerical modeling, which provides a virtual testing ground for hypervelocity impacts. However, such simulations are exceptionally demanding. They must resolve microsecond fracture propagation, long-term dispersal, and fragment interactions. This dual timescale requirement forces tiny integration steps over millions of cycles, making numerical robustness paramount. Any instability in a long-duration run can invalidate fragment statistics (mass, size, velocity) that support probabilistic debris-environment models and design guidelines.
 
\medskip\noindent
Computational methods for fracture can be categorized into mesh-free and mesh-based approaches. Mesh-free methods such as \acrfull{SPH}, validated against hyper-velocity experiments \citep{remington2020numerical,giannaros2019hypervelocity}, handle extreme deformations and mass "splashing" (i.e., the rapid, fluid‐like ejection of discrete material particles under high strain‐rate conditions) naturally. Mesh-based strategies fall into two main categories: damage-based formulations (e.g., phase-field, Lip-field) \citep{francfort1998revisiting,miehe2010phase,chevaugeon2022lipschitz} model cracks with a scalar damage field that degrades stiffness via energy- or strain-driven evolution laws. Phase-field methods capture crack initiation, propagation, branching, and coalescence by solving an additional variational partial differential equation (PDE) over a narrow diffusive zone; they require fine meshes and the solution of coupled systems of equations. On the other hand, sharp crack approaches explicitly represent crack surfaces using enrichment methods like \acrfull{XFEM} \citep{moes1999finite} or \acrfull{CZM} \citep{dugdale1960yielding,barenblatt1962mathematical,xu1994numerical}. The \acrshort{CZM} inserts zero-thickness interface elements along the continuum's element boundaries. Cohesive elements, governed by traction–separation laws, enable the modeling of crack initiation, propagation, and coalescence. The \acrshort{CZM} is classified as \emph{intrinsic} \citep{xu1994numerical}, when elements are pre-inserted along known crack paths, however introducing an initial elastic regime that can artificially soften the global response and increase sensitivity to mesh resolution \citep{tijssens2000numerical}, or \emph{extrinsic} \citep{camacho1996computational}, when elements are adaptively inserted once a local fracture criterion is met--preserving intact stiffness pre-fracture but remaining mesh dependent without mesh adaptation strategies \citep{seagraves2009advances}. Despite these limitations, the \acrshort{CZM} provides an explicit representation of crack surfaces, enabling fragment separation, precise tracking of fragment kinematics, and robust enforcement of post-fracture contact. These aspects make this method particularly well-suited to dynamic fragmentation scenarios \citep{camacho1996computational,ortiz1999finite,repetto2000finite,zhou2005cohesive,pandolfi1999finite}. Their robustness and scalability have been demonstrated in large-scale 3D fragmentation problems \citep{vocialta20173d,vocialta2018numerical}.

\medskip\noindent
Beyond crack modeling, accurately capturing contact and impact among crack faces and fragments is critical. Though contact mechanics are inherently nonsmooth, they can be regularized. For example, smooth, penalty-based methods use differentiable traction–penetration laws, fit naturally with explicit time-stepping, e.g., \acrfull{CD}, add no extra \acrfull{DOF}, and enforce contact approximately; however, stiff contact penalties severely restrict the time step size \citep{wriggers2006computational, pundir2021coupling, belytschko1991contact}. Lagrange-multiplier formulations (augmented-Lagrangian/mortar \citep{wriggers2006computational, simo1992augmented, wohlmuth2000mortar}) impose the Signorini gap constraint at the configuration level via additional multipliers. They are typically solved implicitly as a coupled system. In contrast, nonsmooth complementarity-based time-stepping enforces these constraints at the velocity level, admits instantaneous velocity jumps at impact, and solves a \acrfull{MLCP} each step to obtain contact impulses and post-impact velocities—thus resolving the intrinsic nonsmoothness of the problem. These methods have demonstrated robustness for systems with many simultaneous collisions (e.g., granular assemblies) \citep{dubois2018contact}, but they lead to significant computational overhead and convergence challenges \citep{acary2008numerical}. Based on these considerations, we adopt a penalty-based contact formulation for its high scalability for large-scale fragmentation simulations.

\medskip\noindent
These methodological choices need to balance scalability and stability. Fully implicit solvers, while unconditionally stable, rapidly become prohibitive in both memory and CPU cost. By contrast, the explicit \acrshort{CD} method (i.e., the \(\beta=0\), \(\gamma=\tfrac12\) Newmark-\(\beta\) scheme) is symplectic, second-order accurate in time, and exhibits excellent long-term energy conservation; properties that make it particularly attractive for large-scale fragmentation studies \citep{kane2000variational}. Yet, it comes with a stability condition on the time step size, and coupling an extrinsic \acrshort{CZM} with a penalty contact law in this explicit scheme triggers several instability mechanisms. First, as cohesive damage approaches zero, cohesive stiffness diverges, causing the stable time step \(\Delta t_c\) to shrink towards zero \citep{vocialta20173d}. Second, a large penalty stiffness used to enforce impenetrability similarly reduces the critical time step \(\Delta t_c\) and can excite high-frequency oscillations in the contact forces. Over the years, various techniques have been proposed to mitigate these instabilities. A rate-dependent cohesive traction law is introduced in \citep{vocialta20173d}. For contact, remedies include bi-penalty formulations \citep{hetherington2012new, hetherington2013bipenalty, kolman2021bi}, mass redistribution schemes \citep{khenous2008mass} and singular-mass adjustments \citep{renard2010singular}. Other efforts focus on the time-integration algorithm itself \citep{deuflhard2008contact, wu2009variational, doyen2011time}. More recently, a fully nonsmooth, energy-consistent formulation coupling extrinsic cohesive elements with unilateral contact has been proposed by \citep{collins2022formulation}. However, its performance on large fragmentation problems remains to be demonstrated.

\medskip\noindent
In this paper, we focus on the long‐term behavior of the explicit CZM–penalty contact coupling. While the engineering community has largely overlooked the origin of this instability, our goal is to identify each potential source and isolate it in simple systems, thereby quantifying its relative contribution to global instability. These diagnostics enable us to identify the root causes of unphysical energy drift and provide a basis for informing mitigation strategy choices. The remainder of the paper is organized as follows. Section \ref{sec:numerical-framework} presents the numerical framework and physical model, detailing the extrinsic \acrshort{CZM} formulation, the penalty-based contact treatment, stability checks, and a representative fragmentation scenario. It also introduces the observed unstable behavior via this representative simulation. Section \ref{sec:instability-mechanisms} breaks the instability into three potential sources and quantifies them. In Section \ref{sec:discussion}, we discuss the broader implications of our findings, assess a different contact-penalty formulation--where the penalty stiffness is dynamically adjusted based on local cohesive interface conditions--and discuss potential solutions. Finally, Section \ref{sec:conclusion} concludes and gives directions for future work.
\section{Numerical framework}
\label{sec:numerical-framework}

\noindent
Dynamic fragmentation of brittle solids demands methods that (1) preserve stiffness until failure, (2) resolve crack nucleation, propagation, branching, and coalescence, (3) capture stress‐wave interactions accurately, (4) robustly handle contact and impact between crack faces and fragments, and (5) maintain stability in long simulations without losing scalability. In this work, we employ an extrinsic \acrlong{CZM} with a linear \acrfull{TSL}. We first present the governing equations and discretization of the \acrshort{CZM} and contact formulation, followed by the explicit time-integration scheme and associated stability considerations. We then describe the problem setup, including mesh properties, material parameters, loading conditions, and outputs of interest. Finally, a severe instability is presented using a representative simulation. This framework has been implemented and is studied within the open-source code \texttt{Akantu} \citep{richart2024akantu}.

\subsection{Cohesive-zone model}
\label{ssec:cohesive-zone-model}
\begin{figure}[h!]
  \centering
  \includegraphics[scale=1.]{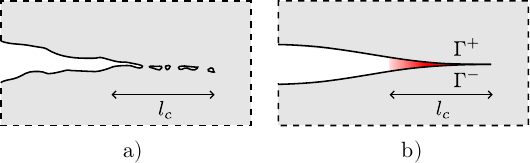}
  \caption{
    \textbf{(a)} Initiation, propagation, and coalescence of microcracks in the fracture process zone \(l_c\), ahead of the main crack tip.  
    \textbf{(b)} Idealized version of the crack where all inelastic processes are represented via a cohesive zone model. \(\Gamma^+\) represents the upper surface of the crack and \(\Gamma^-\) the lower surface.
  }
  \label{fig:cohesive-zone}
\end{figure}

The \acrshort{CZM} regularizes fracture by replacing the sharp crack tip singularity with a finite process zone of characteristic length \(l_c\) ahead of the crack tip (Figure \ref{fig:cohesive-zone}). All inelastic processes within that region are confined to cohesive surfaces  \(\Gamma\) where microcracks initiate, propagate, and coalesce.

\noindent
Denoting the displacement fields on the opposing faces \(\Gamma^+\) and \(\Gamma^-\) by \(\mathbf{u}^+\) and \(\mathbf{u}^-\), the jump
\begin{equation}
    \boldsymbol{\delta} = \boldsymbol{u}^+ - \boldsymbol{u}^-,
\end{equation}
decomposes in the local orthonormal frame \((\boldsymbol{n},\boldsymbol{t})\) of the crack mid-plane as
\begin{equation}
  \delta_n = \boldsymbol\delta\cdot\boldsymbol n,
  \quad
  \delta_t = \boldsymbol\delta\cdot\boldsymbol t.
  \label{eq:delta_decomposition}
\end{equation}
The traction vector applied to the corresponding free surfaces, thereby keeping the crack faces closed, derives from a potential \(\phi(\boldsymbol{\delta})\),
\begin{equation}
    \mathbf{T}(\boldsymbol{\delta}) = \frac{\partial\phi}{\partial\boldsymbol\delta}.
    \label{eq:traction_potential}
\end{equation}

\subsubsection{Cohesive elements}
To discretize \acrshort{CZM} in a \acrfull{FEM} framework, zero‐thickness cohesive elements are placed along the faces of neighboring continuum elements. Two distinct formulations of the traction–separation potential \(\phi\) give rise to different insertion strategies:

\begin{itemize}
    \item \textbf{Intrinsic CZM:} Cohesive elements are pre‐inserted along predefined paths (e.g., grain boundaries or interfaces) \citep{xu1994numerical}. The associated \acrshort{TSL} includes an initial elastic regime. This addition of compliance can modify the effective stiffness of the structure, even in uncracked regions, and increase sensitivity to mesh resolution. Intrinsic models are, however, straightforward to implement and naturally suited to fracture along known paths.

    \item \textbf{Extrinsic CZM:} Cohesive elements are inserted dynamically when a local fracture criterion is met \citep{camacho1996computational, pandolfi1999finite}. This approach preserves bulk stiffness in intact regions and supports arbitrary crack paths, making it well-suited for dynamic fragmentation. However, topological changes in the mesh at insertion make the implementation more challenging.
\end{itemize}
In this work, we adopt the extrinsic approach for its benefits: cohesive element insertion occurs only when and where needed, thereby minimizing computational cost.

\subsubsection{Traction–Separation Law}

\begin{figure*}[h!]
    \centering
    \includegraphics[scale=1.]{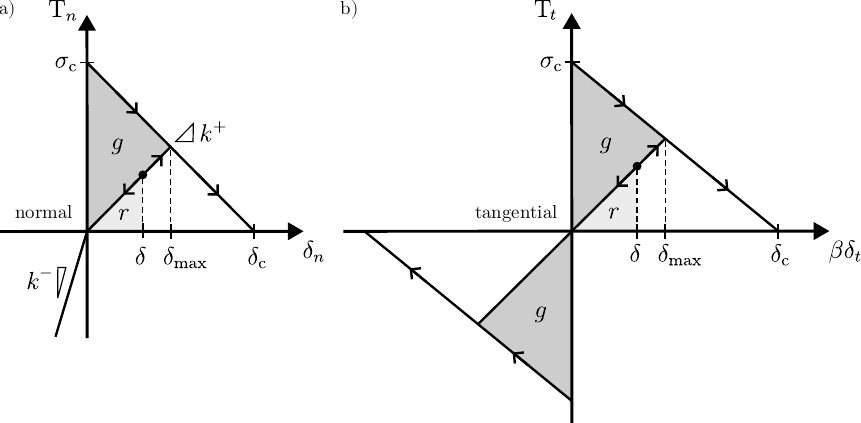}
    \caption{
        Linear irreversible \acrshort{TSL} for extrinsic cohesive elements, following the formulation of Camacho and Ortiz \citep{camacho1996computational}, with a penalty-based contact enforcement. \textbf{(a)} Normal component. \textbf{(b)} Tangential component.
    }
    \label{fig:TSL_Camacho_Ortiz}
\end{figure*}

We adopt the linear irreversible \acrshort{TSL} of Camacho and Ortiz \citep{camacho1996computational} presented in Figure \ref{fig:TSL_Camacho_Ortiz} (\citep{ortiz1999finite} in 3D). This law derives from the cohesive potential energy per unit area,
\begin{equation}
    \phi(\delta) = \frac{1}{2}\sigma_c\delta\left(2-\frac{\delta}{\delta_c}\right),
\end{equation}
where \(\sigma_c\) represents the local strength of the material, and \(\delta_c\) is the critical opening beyond which complete decohesion occurs. The fracture toughness per unit area is defined by \(G_c=\sigma_c\delta_c/2\). Instead of relying on the relative displacement vector \(\boldsymbol\delta\), the law introduces an effective displacement,
\begin{equation}
    \delta = \sqrt{\delta_n^2 + \beta^2\delta_t^2},
\end{equation} 
weighing normal and tangential contributions via the dimensionless parameter \(\beta\). The traction vector \(\mathbf{T}\) is derived from the potential (Eq.~\ref{eq:traction_potential}) as follows:
\begin{equation}
    \mathbf{T} = \frac{\partial\phi}{\partial\boldsymbol{\delta}}=\frac{\rm{T}(\delta)}{\delta}\left(\beta^2\delta_t\boldsymbol t + \delta_n\boldsymbol n\right),
    \quad
    \rm{T}(\delta) = \sigma_c\left(1-\frac{\delta}{\delta_c}\right).
\end{equation}
Let \(\delta_{\max}\) be the maximum effective opening. Since unloading/reloading is linear elastic, the secant cohesive stiffness \(k^+\) at any \(\delta\le\delta_{\max}\) is defined by
\begin{equation}
    k^+ = \frac{\rm{T}(\delta_{\max})}{\delta_{\max}} = \frac{\sigma_c}{\delta_\text{max}}\left(1-\frac{\delta_\text{max}}{\delta_c}\right),
    \label{eq:k_plus_dmax}
\end{equation}
and decreases irreversibly with \(\delta_{\max}\). During the decohesion process, the energy is partitioned into reversible cohesive energy $r$ and irreversible fracture energy $g$ associated with surface creation. These contributions per unit area are defined as follows:
\begin{equation}
    r = \frac{1}{2}k^+\delta^2, \quad
    g = \frac{1}{2}\sigma_c\delta_\text{max}.
    \label{eq:cohesive_fracture_energy}
\end{equation}

\subsubsection{Dynamic insertion criterion}
\label{ssec:dynamic-insertion-criterion}
The dynamic insertion of a cohesive element takes place on a facet of the mesh when an effective driving stress $\sigma_\text{eff}$ reaches the local cohesive strength $\sigma_c$. This effective stress, introduced in \citep{camacho1996computational}, combines the normal and tangential stress components, with their relative contribution controlled by the same dimensionless parameter $\beta$:
\begin{equation}
  \sigma_{\rm eff}
  = \sqrt{\sigma_n^2 + \frac{\sigma_t^2}{\beta^2}}\,.
\end{equation}
Since no fracture should happen in compression, \(\sigma_n<0\) is taken as zero. When \(\sigma_{\rm eff}\ge\sigma_c\), a cohesive element is dynamically inserted along that facet by locally modifying the mesh topology. The node-splitting technique is implemented in \texttt{Akantu}, and adapted for High-Performance Computing (HPC) use \citep{vocialta20173d, richart2024akantu}.  

\subsection{Penalty-based contact enforcement}
\label{ssec:penalty-based-contact}
\noindent
Up to now, the \acrshort{TSL} has been based on an effective displacement that is sign‐independent—i.e.\ the \(\rm{T}\)–\(\delta\) curves are symmetric about the origin. While this is acceptable for tangential sliding, it fails under normal compression (\(\delta_n<0\)), where the two surfaces would unrealistically interpenetrate. To penalize and limit such interpenetration, we embed a simple penalty‐based contact law into our cohesive formulation \citep{wriggers2006computational,pundir2021coupling} (Figure \ref{fig:TSL_Camacho_Ortiz}a),  

\begin{equation}
    \text{T}_n = k^-\delta_n, \quad \delta_n < 0, \quad k^->0,
\end{equation}
so that the contact penalty \(k^-\to\infty\) would recover exact impenetrability \citep{kikuchi1988contact}. This penalty can be interpreted as a spring stiffness, and the resulting reversible contact energy per unit area yields
\begin{equation}
    c = \frac{1}{2}k^-\delta_n^2, \quad \delta_n < 0. 
    \label{eq:contact_energy}
\end{equation}

\paragraph{Penalty‐parameter selection.} 
The choice of \(k^-\) is non-trivial \citep{wriggers2006computational}. It must balance two competing needs: 
\begin{itemize}
  \item \textbf{Accuracy:} \(k^-\) must be large enough that typical interpenetrations remain below a prescribed tolerance. This tolerance is typically chosen as a fraction of the continuum element size \(h\). Note that this choice is therefore dependent on the loading conditions and the material properties. 
  \item \textbf{Stability:} Excessively large \(k^-\) leads to ill‐conditioned stiffness matrices and, in explicit dynamics, to prohibitively small time-steps.
\end{itemize}
A common guideline \citep{wriggers2006computational} is to scale the normal penalty with the bulk stiffness over an element representative length \(h\):
\begin{equation}
    k^- \approx \alpha\,\frac{E}{h},
    \quad
    \alpha\sim 10^1\text{–}10^4,
    \label{eq:penalty_stiffness}
\end{equation}
where \(E\) is Young’s modulus. 

\subsection{Time integration and stability bound}
\label{ssec:time-integration}

\noindent
We use the explicit \acrfull{CD} scheme (Newmark–\(\beta\) with \(\beta=0\), \(\gamma=\tfrac12\)), which is second‐order accurate, preserves energy in a weak sense and when applied to Hamiltonian systems, is symplectic, i.e.\ it preserves the Hamiltonian structure of the equations of motion \citep{kane2000variational}. In the absence of damping, it advances the semi‐discrete equations of motion
\begin{equation}
  \mathbf{M}\,\mathbf{\ddot u} + \mathbf{K}\,\mathbf{u} = \mathbf{f}^{\rm ext}, 
  \label{eq:semi-discrete-motion}
\end{equation}
from step \(n\) to \(n+1\) (separated by the time step \(\Delta t\)) via 
\begin{align}
  \mathbf{u}_{n+1} &= \mathbf{u}_n + \Delta t\,\mathbf{v}_n + \tfrac12\,\Delta t^2\,\mathbf{a}_n
    &&\text{(update displacement)}\label{eq:central-difference-u}\\
  \mathbf{a}_{n+1} &= \mathbf{M}^{-1}\bigl(\mathbf{f}^{\rm ext}_{n+1} - \mathbf{K}\,\mathbf{u}_{n+1}\bigr)
    &&\text{(compute acceleration)}\label{eq:central-difference-v}\\
  \mathbf{v}_{n+1} &= \mathbf{v}_n + \tfrac12\,\Delta t\,\bigl(\mathbf{a}_n + \mathbf{a}_{n+1}\bigr)
    &&\text{(update velocity)}
    \label{eq:central-difference-a}
\end{align}
where \(\mathbf{v}=\mathbf{\dot u}\), \(\mathbf{a}=\mathbf{\ddot u}\), \(\mathbf{M}\) is the (lumped) mass matrix, \(\mathbf{K}\) the assembled stiffness matrix (bulk + cohesive), and \(\mathbf{f}^{\rm ext}\) the external forces.  

\medskip\noindent
As shown by the well-known spectral‐radius analysis in Appendix~\ref{app:cd_stability}, we recall that the explicit \acrshort{CD} integrator is conditionally stable \citep{belytschko2014nonlinear}. To study this stability limit, we write the discrete state update as
\[
  \mathbf{x}_{n+1} = \mathbf{\Phi}\,\mathbf{x}_n,
\]
with the state \(\mathbf{x}=[\mathbf{u}, \mathbf{v}]^\top\) and where \(\mathbf{\Phi}\) is the state‐transition matrix, relating to consecutive steps $n$ and $n+1$. Stability requires the spectral radius $\rho$ of $\mathbf{\Phi}$ (i.e., its largest absolute eigenvalue) to satisfy \(\rho(\mathbf{\Phi})\le1\). Rewriting this condition is equivalent to choosing a time step
\begin{equation}
    \Delta t \;\le\; \Delta t_c
    \;=\;
    \frac{2}{\sqrt{\lambda_{\max}(\mathbf{M}^{-1}\mathbf{K})}}
    \;=\;
    \frac{2}{\omega_{\max}},
    \label{eq:critical_time_step}
\end{equation}
where \(\omega_{\max}=\sqrt{\lambda_{\max}(\mathbf{M}^{-1}\mathbf{K})}\) is the highest natural frequency of the undamped system.

\medskip\noindent
Computing the exact stability limit requires a complete eigenvalue analysis of the combined mass–stiffness matrix, which is prohibitively expensive for large-scale fragmentation simulations and evolves constantly due to topological changes of the mesh. Following the work of \citep{osti_5429002}, we can apply Gershgorin’s circle theorem to bound the largest eigenvalue of \(\mathbf{M}^{-1}\mathbf{K}\):
\[
  \lambda_{\max}(\mathbf{M}^{-1}\mathbf{K})
  \;\le\;
  \max_i\frac{K_{ii} + \sum_{j\neq i} |K_{ij}|}{M_{ii}}.
\]
In \citep{osti_5429002} impact-only context, each diagonal entry \(K_{ii}\) includes bulk stiffness contributions at \acrfull{DOF} \(i\) plus penalty contributions from facets already in contact. By contrast, in a cohesive‐zone model, new contact faces can nucleate anywhere, so we conservatively include \emph{all} potential cohesive/contact facets:
\[
  K_{ii}
  = \sum_{e\ni i}\int_{\Omega^e}\! \mathbf{B}^T\mathbf{D}\,\mathbf{B}\,dV
    \;+\; \sum_{\Gamma\ni i} k^-\,A_{\Gamma},
\]
where \(\mathbf{B}\) is the standard element strain–displacement matrix and \(\mathbf{D}\) is the constitutive matrix mapping strains to stresses; the first sum runs over every bulk element \(e\) supporting \acrshort{DOF} \(i\), and the second over every facet \(\Gamma\) whose cohesive law involves \acrshort{DOF} \(i\) (with facet area \(A_{\Gamma}\)). It follows that
\begin{equation}
  \Delta t \;\le\; 2\min_i
    \sqrt{\frac{M_{ii}}{\,K_{ii} + \sum_{j\neq i}|K_{ij}|\,}}
  \,,
  \label{eq:gershgorin-penalty}
\end{equation}
so larger penalty parameters \(k^-\) increase \(K_{ii}\) and directly shrink \(\Delta t\), reflecting the trade‐off between contact stiffness and allowable time‐step size. By contrast, the standard Courant–Friedrichs–Lewy (CFL) condition
\citep{courant1928partiellen}
\begin{equation}
    \Delta t \;\le\;\frac{r_{\min}}{c},
    \quad
    r_{\min} = \min_i\{r_i\}, \quad \text{ with } r_i \text{ the characteristic length of element }i\text{ in the mesh,}
    \label{eq:cfl_bound}
\end{equation}
It depends only on the bulk wave speed \(c\) and a representative element length \(r_{\min}\) (e.g., inradius), neglects all contact penalty stiffening, and can significantly overestimate the stable time-step when penalty‐based contact is active. 

\subsection{Problem setup}
\label{ssec:problem-setup}

\begin{figure}[h!]
    \centering
    \includegraphics[scale=1]{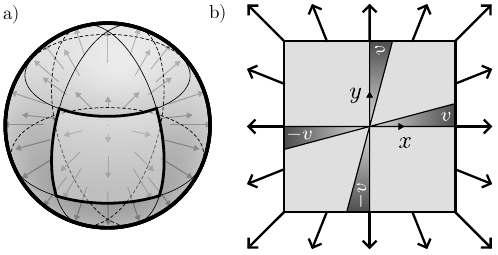}
    \caption{
        \textbf{(a)} Physical system: expanding hollow sphere.  
        \textbf{(b)} Planar analog: square plate of side \(L\), with velocity field \(\textbf{v}(r)=\dot\varepsilon\,\textbf{r}\).
    }
    \label{fig:geometry_2D_equivalent}
\end{figure}

We simulate the dynamic fragmentation of an exploding hollow sphere via its two-dimensional analog, a square plate of side length \(L=10^{-2}\,\mathrm{m}\) (Figure~\ref{fig:geometry_2D_equivalent}). Uniform radial expansion is imposed by initializing every node velocity, following a radial velocity field
\begin{equation}
  \mathbf{v}(r, t=0) \;=\; \dot\varepsilon\,\mathbf{r},
\end{equation}

\noindent
The applied loading defines a spatially uniform strain rate $\dot\varepsilon$. While unloading stress waves emanate from the plate edges from $t=0$, we choose $\dot\varepsilon$ high enough that the central region of the plate experiences an approximately uniform fragmentation, before the waves reach it. We normalize the strain rate by the characteristic strain rate \(\dot\varepsilon_0\) of the material, defined by \citep{drugan2001dynamic} as
\begin{equation}
    \dot\varepsilon_0 = \frac{\sigma_c}{Et_0},
    \quad
    t_0 = \frac{EG_c}{\sigma_c^2c},
    \quad
    c = \sqrt{\frac{E}{\rho(1-\nu^2)}},
\end{equation}
with \(c\) the bulk wave speed in plane-stress conditions. The characteristic time \(t_0\) is the time needed by the cohesive element to reach its maximum opening \(\delta_c\) when subjected to \(\dot\varepsilon_0\) \citep{camacho1996computational}. From the material properties defined in Section~\ref{sssec:material_properties}, the characteristic strain rate is \(\dot\varepsilon_0\approx2.6\times10^4 \mathrm{ s}^{-1}\) and we choose a normalized strain rate of \(\hat{\dot\varepsilon} = \dot\varepsilon/\dot\varepsilon_0 = 0.2\). 

\subsubsection{Material properties}
\label{sssec:material_properties}
We perform our simulations using AD‑995 alumina; its key properties are summarized in Table~\ref{tbl:material}. To capture the scatter in local failure strength due to microstructural flaws, each facet’s cohesive stress \(\sigma_c\) is sampled from a Weibull distribution \citep{weibull1939statistical}. Specifically, we employ the two‑parameter form
\begin{equation}
    F(\sigma_c)=1-\exp\left(\left(\frac{\sigma_c}{\lambda}\right)^m\right),
\end{equation}
where \(m\) denotes the Weibull modulus (shape) and \(\lambda\) is the scale parameter. This statistical description of strength variability has been studied in depth in the context of dynamic fragmentation \citep{levy2010dynamic, levy2012dynamic}.

\begin{table}[h!]
  \centering
  \begin{tabular}{l c}
    \toprule
    \textbf{Parameter} & \textbf{Value} \\
    \midrule
    Young’s modulus, \(E\) [GPa]                      & 370              \\
    Density, \(\rho\) [kg/m\(^3\)]                    & 3900             \\
    Cohesive strength distribution, $\sigma_c$        & Weibull          \\
    \quad Scale parameter, \(\lambda\) [MPa]          & 262              \\
    \quad Shape (Weibull modulus), \(m\)              & 10               \\
    Fracture toughness, \(G_c\) [J/m\(^2\)]           & 50               \\
    Poisson’s ratio, \(\nu\)                          & 0.22             \\
    \bottomrule
  \end{tabular}
  \caption{AD‑995 alumina parameters.  Cohesive strength is drawn from a Weibull distribution \(F(\sigma_c)=1-\exp[-(\sigma_c/\lambda)^m]\), with  \(\lambda=262\) MPa tied to the direct tensile strength and  \(m=10\) chosen from typical alumina scatter.}
  \label{tbl:material}
\end{table}

\subsubsection{Mesh discretization and output metrics}
We present a single high-resolution run in Section~\ref{ssec:observed_instability}. The plate is discretized with linear triangular (T3) elements of uniform side length \(h = 3.7\times10^{-5}\,\mathrm{m}\), yielding three elements in the cohesive zone and approximately \(2\times10^5\) global degrees of freedom, before any cohesive elements insertion.

\medskip\noindent
To quantify fragmentation, we track \emph{energy partitioning} and \emph{fragment statistics}. The bulk elastic and kinetic energies are denoted by $\mathcal{U}$ and $\mathcal{K}$, respectively, and are computed from displacements and velocities, which are stored as nodal values. The recoverable cohesive, contact, and dissipated fracture energy densities, stored at quadrature points, (Eq.~\ref{eq:cohesive_fracture_energy} and \ref{eq:contact_energy}) are integrated and summed over all the cohesive facets, respectively, into $\mathcal{R}$, $\mathcal{C}$, and $\mathcal{G}$. The external work stays null in the studied scenario since no external loading is applied. As a supplementary output metric, we also extract the fragment count \(N_{\rm frag}\).

\medskip\noindent
We summarize below the key signatures that a well‐behaved extrinsic \acrshort{CZM} simulation should exhibit:
\begin{enumerate}
  \item \textbf{Energy conservation.} At all times \(t\), the total energy injected into the system \(\mathcal{E}_{\rm inj}(t)\) should equal the sum of elastic, kinetic, and dissipated contributions, summed into \(\mathcal{E}_{\rm tot}(t)\), the total energy:
  \begin{equation}
      \mathcal{E}_{\rm inj}(t)=\mathcal{E}_{\rm tot}(t)
      \label{eq:energy_balance}
  \end{equation}
  where,
  \begin{equation}
      \mathcal{E}_{\rm tot}(t)=\underbrace{\mathcal{U}(t) + \mathcal{R}(t) + \mathcal{C}(t)}_{\displaystyle \text{ (recoverable elastic)}} + \underbrace{\mathcal{K}(t)}_{\displaystyle \text{ (kinetic)}} + \underbrace{\mathcal{G}(t)}_{\displaystyle \text{ (fracture dissipation)}},
  \end{equation}
  The left‐hand side of Equation~\ref{eq:energy_balance} may also be written as
  \begin{equation}
      \mathcal{E}_{\rm inj}(t) = \mathcal{W}^{\rm ext}(t) + \mathcal{E}_0,
  \end{equation}
  with \(\mathcal{E}_0\) the initial energy present in the system (purely kinetic, due to the initial radial velocity field).  
  \item \textbf{Predictable energy partitioning.} During crack nucleation and growth, bulk elastic and kinetic energies are progressively converted into cohesive and contact energies and dissipated as fracture energy. After fragmentation completes, all recoverable and kinetic energies oscillate around constant means, indicating a stable equilibrium. Dissipated energy reaches a plateau.
  
  \item \textbf{Fragment‐count plateau.} The number of fragments \(N_{\rm frag}(t)\) rises from one to a finite value during the main fragmentation event, and then remains constant.
\end{enumerate}

\subsection{Observed Instability}
\label{ssec:observed_instability}
To verify the idealized behaviors outlined above, we perform the numerical experiment described in Section \ref{ssec:problem-setup}, using the prescribed mesh discretization and material parameters. We set the time step to 
\begin{equation*}
    \Delta t = 0.4\,\Delta t_{c,G}\,,
\end{equation*}
where \(\Delta t_{c,G}\) is the critical time increment estimated via a Gershgorin bound (Eq.~\ref{eq:gershgorin-penalty}) employing a contact‐penalty \(k^-\) of \(10^{1}E/h\) (Eq.~\ref{eq:penalty_stiffness}).

\begin{figure}[h!]
    \centering
    \includegraphics[scale=1.]{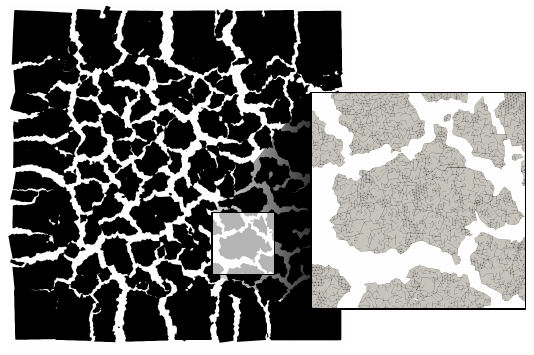}
    \caption{
    Fragmentation snapshot taken at $t=3.75\times10^{-5}\text{ s}$. The magnified view reveals partially damaged cohesive elements within an individual fragment.
    }
    \label{fig:frag2D_snapshot}
\end{figure}

\begin{figure}[h!]
    \centering
    \includegraphics[scale=1]{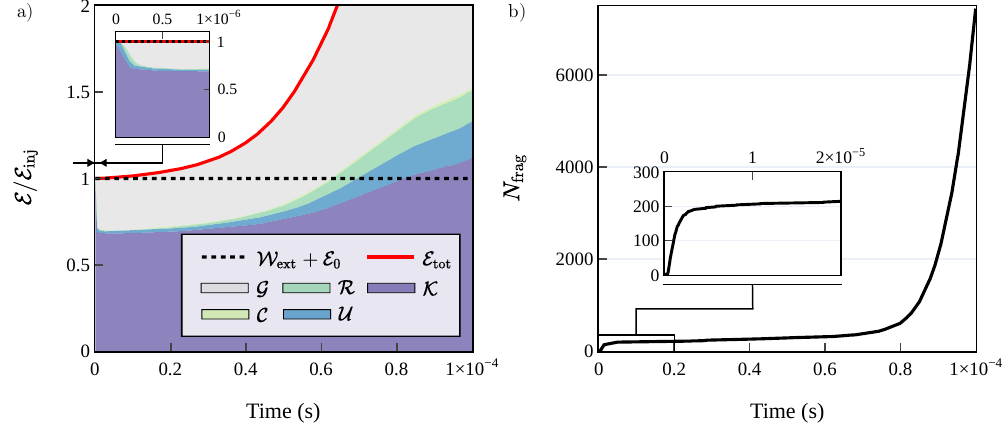}
    \caption{
        \textbf{(a)} Time history of energy components, kinetic \(\mathcal{K}\), fracture \(\mathcal{G}\), and elastic energies, normalized by the total injected energy 
        \(\mathcal{E}_\mathrm{inj} = \mathcal{W}_\mathrm{ext}(t_f) + \mathcal{E}_0\). Rather than settling to steady values after fragmentation, all components exhibit continuous growth, revealing unphysical energy drift.  
        \textbf{(b)} Evolution of the fragment count over time. After an initial plateau, spurious high‐frequency oscillations induce non‐physical additional fragments, underscoring the impact of numerical instabilities on fragment statistics.
    }
    \label{fig:frag2D_instability}
\end{figure}

\medskip\noindent
A snapshot of the resulting fragmentation pattern is visible in Figure~\ref{fig:frag2D_snapshot}, and we present the quantitative results in Figure \ref{fig:frag2D_instability}. They reveal an unexpected instability despite satisfying the stability condition $\Delta t<\Delta t_c$. Panel (a) shows energy partitioning over time with stacked contributions from the various energy components, summed as \(\mathcal{E}_{\rm tot}\) and normalized by the injected energy \(\mathcal{E}_{\rm inj}\). Initially, since kinetic energy is in excess, it is dissipated as fracture energy (see zoomed region). However, after the main fragmentation event, the total energy does not stabilize around a constant value, but instead exhibits a continuous drift above unity. In panel (b), the energy drift manifests as non‐physical fragmentation. After an initial quasi-steady phase at \(N_{\rm frag}\approx 200\), the fragment count grows rapidly and continuously, indicating an unstable and unphysical fragmentation process. Although these results suggest a strong stability issue in our simulations, they do not pinpoint the root cause. In Section~\ref{sec:instability-mechanisms}, we examine three candidate sources of instability.

\section{Sources of instability}
\label{sec:instability-mechanisms}
\begin{figure*}[h!]
    \centering
    \includegraphics[scale=1]{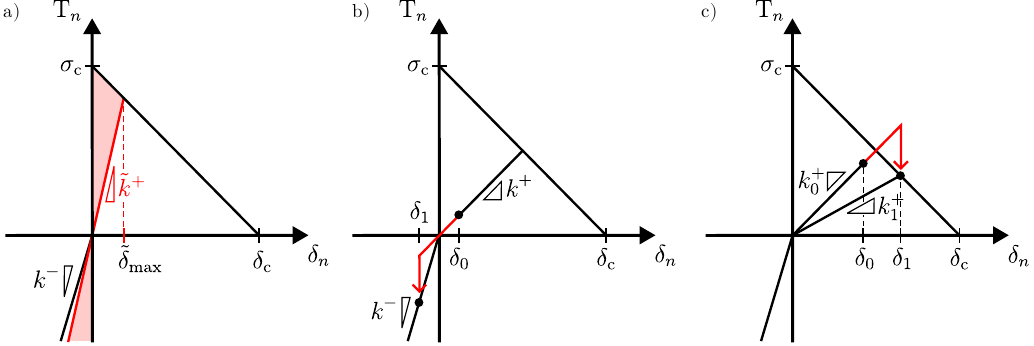}
    \caption{
        Schematics of the traction/separation path of a cohesive element in the normal direction, highlighting three potential sources of numerical instabilities.
        \textbf{(a)}  In the undamaged limit (\(d\to0\), i.e., \(\delta_{\max}\to 0\)), the cohesive stiffness \(k^+\) diverges, shrinking the explicit integrator's critical time-step \(\Delta t_c\). For a fixed time-step, the stiffness threshold \(\tilde{k}\) separates stable from unstable behavior.
        \textbf{(b)} The abrupt switch between cohesive stiffness \(k^+\) and contact penalty \(k^-\) violates the smooth evolution assumption of explicit \acrshort{CD} integration. 
        \textbf{(c)} The onset of softening introduces an additional discontinuity in the traction–separation response, which similarly could lead to numerical errors in explicit integration.
    }
    \label{fig:instabilities_tsl}
\end{figure*}

\noindent
From our test run, we showed that coupling extrinsic \acrshort{CZM} with penalty-based contact under explicit time integration exhibits numerical instabilities that can corrupt the energy balance and fragment count. In this section, we identify and review three potential sources of instability, schematized in Figure~\ref{fig:instabilities_tsl}. We first evaluate the high–initial–stiffness "hot‐spot" effect (Section~\ref{ssec:high_initial_stiffness}), then examine two instances of nonsmooth stiffness transitions (Sections~\ref{ssec:cohesive-contact} and~\ref{ssec:softening}), and quantify each mechanism using idealized and simplified models.

\subsection{High initial stiffness}
\label{ssec:high_initial_stiffness}
When a cohesive element is first inserted, its low damage state yields an extremely large stiffness (Figure~\ref{fig:instabilities_tsl}a). Because explicit \acrshort{CD} time integration requires a time-step \(\Delta t\) small enough to resolve the stiffest springs, such a "hot spot" can transiently violate the stability bound and inject spurious energy into the system.

\begin{figure}[H]
    \centering
    \includegraphics[scale=1]{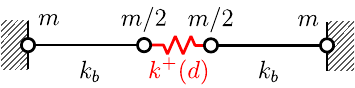}
    \caption{Schematic of a 1D system made of two bulk elements connected by a cohesive element. The bulk elements have stiffness $k_b$ and the cohesive element has stiffness $k^+(d)$ as in Eq.~\eqref{eq:k_plus_damage}. Both ends are fixed.}
    \label{fig:bulk_cohesive_sys}
\end{figure}

\paragraph{Idealized model.}
To quantify this effect, we consider the one-dimensional spring–mass chain in Fig.~\ref{fig:bulk_cohesive_sys}. Two identical bulk springs of stiffness $k_b$ are fixed at their ends and connected by a damage-dependent cohesive spring whose stiffness is
\begin{equation}
  k^{+}(d) = \frac{\sigma_{c}}{\delta_{c}}\,\frac{1 - d}{d},\qquad
  d = \frac{\delta_{\max}}{\delta_{c}} \in (0,1], \qquad \text{(Eq.~\ref{eq:k_plus_dmax})},
  \label{eq:k_plus_damage}
\end{equation}
which diverges as \(d\to0\).  A direct application of the explicit \acrshort{CD} stability bound (Eq.~\ref{eq:critical_time_step}) on the system yields
\begin{equation}
    \Delta t \leq \Delta t_\text{c}(k^+, k_b, m) = \frac{\sqrt{2m}}{\sqrt{2k^+(d) + k_b}}.
    \label{eq:critical_time_step_2}
\end{equation}
It shows that, for any fixed $\Delta t$, there exists a maximum admissible cohesive stiffness $\tilde k^+(\Delta t)$ (or equivalently a minimum damage threshold $\tilde d(\Delta t)$) below which the scheme is unstable.

\begin{figure*}[h!]
    \centering
    \includegraphics[scale=1]{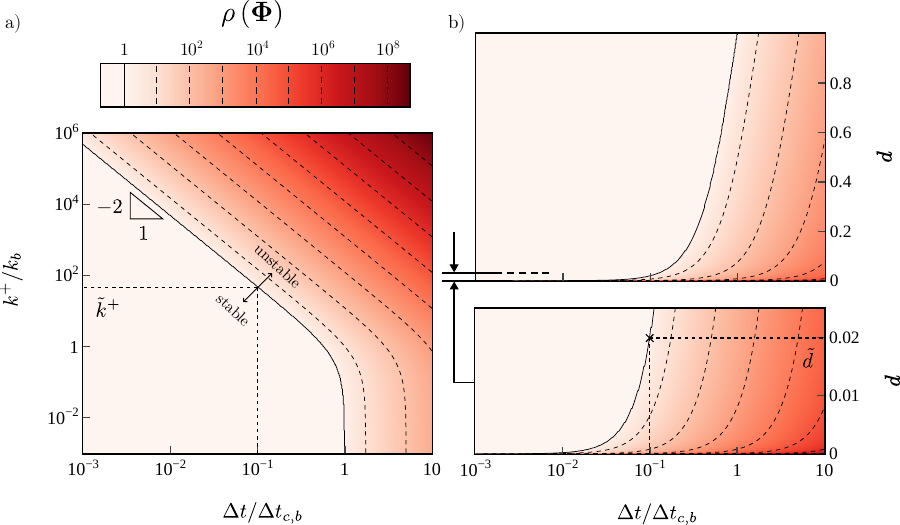}
    \caption{
    Spectral-radius stability analysis for the discrete state-transition matrix $\mathbf{\Phi}$ of the mass–spring–mass model (Fig.~\ref{fig:bulk_cohesive_sys}). The per-step amplification factor is $\rho(\mathbf{\Phi})=\max_i|\lambda_i(\mathbf{\Phi})|$; stability requires $\rho\le1$. \textbf{(a)} $\rho$ in the $(\Delta t, k^+)$ with $\Delta t$ normalized by $\Delta t_{c,b}$ and $k^+$ normalized by $k_b$. \textbf{(b)} $\rho$ in the $(\Delta t,d)$-plane via Eq.~\eqref{eq:k_plus_damage}. \textbf{(c)} Enlarged view of (b); the contour $\rho=1$ defines $\tilde d(\Delta t)$ (e.g., at $\Delta t=0.1\,\Delta t_{c,b}$, instability arises for $d<\tilde d\approx0.02$ or equivalently, $k^+>\tilde k^+\approx50k_b$).}
    \label{fig:low_damage}
\end{figure*}

\medskip\noindent
While choosing this time step is ideal, it should be updated as the system evolves. An arbitrarily large cohesive stiffness $k^+$ would, in turn, reduce $\Delta t$ to an arbitrarily small value. In practice, the time step in explicit dynamics is usually determined by the standard CFL condition, which depends only on the bulk stiffness and neglects cohesive contributions. To follow standard practice and to assess the impact of cohesion on stability, we therefore fix the critical time step at the bulk-only limit.
\begin{equation}
  \Delta t_{c,b}
  \;=\;
  2\sqrt{\frac{m}{k_b}}
  \quad\text{(Eq.~\ref{eq:cfl_bound})},
  \label{eq:bulk_critical_time-step}
\end{equation}
Following Appendix~\ref{app:cd_stability}, we form the discrete state-transition matrix $\mathbf{\Phi}$ for a normalized test case ($k_b=1$, $m=\tfrac12$, $\sigma_c=\delta_c=1$) and compute its spectral radius $\rho(\mathbf{\Phi})$. Figure~\ref{fig:low_damage} presents the values of $\rho$ as a function of the time step $\Delta t$ and the cohesive stiffness $k^+$ or damage $d$. Panel a) presents the results on a log–log scale: the light-shaded region corresponds to the stability domain \(\rho \leq 1\), while the contour \(\rho = 1\) marks the stability limit, where the cohesive stiffness \(k^+\) reaches its maximum admissible value \(\tilde{k}^+\) for a fixed time step. When $k^+\ll k_b$, the stability limit approaches the bulk-only value $\Delta t_{c,b}$. Conversely, for $k^+\gg k_b$, it converges logically to the cohesive-only limit $\Delta t_c^+$ (Eq~\ref{eq:bulk_critical_time-step} with $k^+$). Standard practice relies on multiplying the critical time step by a safety factor; for example, with $\Delta t=0.1\Delta t_{c,b}$, the system becomes unstable at $\tilde k^+ = 50 k_b$, as indicated by the visual guidelines. Figure~\ref{fig:low_damage}b replots $\rho$ as a function of the time step and substitutes $k^+$ by cohesive damage $d$ using Eq.~\eqref{eq:k_plus_damage}. The contour \(\rho=1\) now defines the damage threshold $\tilde d$ for a fixed time step. Figure~\ref{fig:low_damage}c shows a close‐up of panel~\ref{fig:low_damage}b: at $\Delta t=0.1\,\Delta t_{c,b}$ and for the chosen material properties, instability occurs when \(d<\tilde d\approx0.02\).

\medskip\noindent
The spectral-radius analysis implies that for any fixed time step, there exists a bound that delimits stability from instability. Under complex loading scenarios, newly inserted cohesive elements sample a wide range of initial damage, and, if the time step is fixed, some will unavoidably fall below the threshold. A practical recommendation is \emph{local monitoring}: flag and track "hot spots" with $d<\tilde d$, or equivalently $k^+>\tilde k^+$, and verify that they soften past the threshold before oscillations propagate to the surrounding bulk. During the test run of Section~\ref{ssec:problem-setup}, fortunately, no cohesive element ever crossed this instability threshold. However, this particular case is due to the choice of a specific loading scenario, characterized by a uniform and high strain rate, which causes cohesive elements to overshoot past the unstable region in one time step. It does not imply that this source of instability can be universally discounted, especially under lower and non-uniform strain rates. Still, it indicates that this mechanism is not the root cause of the observed global instability of Section~\ref{ssec:problem-setup}. 

\medskip\noindent
A qualitatively different mechanism remains: partially damaged cohesive interfaces that close and reopen under oscillations and vibrations of the surrounding bulk. In what follows, we focus on this \emph{cohesive–contact transition} and quantify its effect on the energy balance.

\subsection{Cohesive/contact transition}
\label{ssec:cohesive-contact}

\noindent
After fragmentation, the system contains a dense network of cohesive elements, a lot of them remaining only partially damaged (Figure~\ref{fig:frag2D_snapshot}). Under the influence of stress waves, these interfaces transition repeatedly between cohesion and contact, making their normal opening $\delta$ oscillate around zero. The local response thus switches abruptly between $k^+$ and $k^-$. This problem relates mainly to the chosen contact penalty formulation. Indeed, while the explicit \acrshort{CD} scheme is symplectic (and preserves energy in a weak sense) under the assumption of a smooth evolution of the system~\citep{kane2000variational}, such a discontinuity in the \acrshort{TSL} introduces nonsmoothness that results in energy artifacts, and possibly instability.

\begin{figure}[h!]
    \centering
    \includegraphics[scale=1]{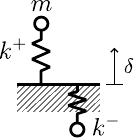}
    \caption{Single‐degree‐of‐freedom (SDOF) system with mass $m$, cohesive stiffness $k^+$ for $\delta>0$, and contact penalty stiffness $k^-$ for $\delta<0$.}
    \label{fig:sdof_system}
\end{figure}

\paragraph{Idealized model and parameterization.}
To isolate and quantify the effect of these transitions, we consider an idealized \acrfull{SDOF} system. A mass~$m$ is attached to a piecewise‐linear spring: when $\delta>0$, the spring stiffness is $k^+$ (cohesion) and when $\delta<0$, it is $k^-$ (contact) (Figure~\ref{fig:sdof_system}). We parameterize any single stiffness change between two consecutive steps~$0\!\to\!1$ by the \emph{oriented stiffness ratio}
\begin{equation}
  \alpha \;:=\; \frac{k_1}{k_0},
\end{equation}
so that a contact$\to$cohesive transition yields $\alpha = k^+/k^-$ and a cohesive$\to$contact one yields $\alpha = k^-/k^+$. This minimal setup captures the essential dynamics of a partially damaged cohesive interface oscillating around $\delta=0$. Within this framework, we first derive the energy error for a single transition, as a function of the time step~$\Delta t$ and the stiffness ratio~$\alpha$ (Section~\ref{sssec:one_step_energy_jump}) and then assess the long‐term stability of the same \acrshort{SDOF} system under repeated stiffness transitions, to determine whether successive energy errors can accumulate into a global instability, as a function of the time step~$\Delta t$ and the reference stiffness ratio $\alpha_{\mathrm{ref}}$ (Section~\ref{sssec:phase_space_stability}).

\subsubsection{Energy error for a single transition}
\label{sssec:one_step_energy_jump}

\begin{figure}[H]
    \centering
    \includegraphics[scale=1]{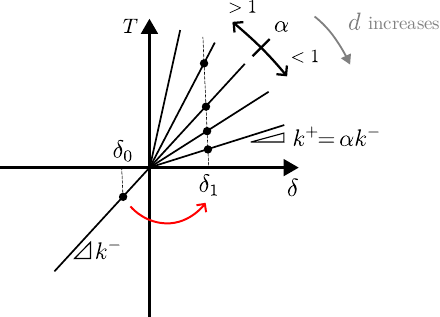}
    \caption{
        Schematic of the traction/separation path for a contact\(\to\)cohesive transition parameterized by the ratio $\alpha=k^+/k^-$.
    }
    \label{fig:sdof_overshoot_schematic}
\end{figure}

We focus on a generic stiffness change from \(k_0 \to k_1\) in the \acrshort{SDOF} model. The initial state is given by
\[
  \mathbf{x}_0 = [\delta_0,\,\dot\delta_0]^\top,
\]
with $\delta_0$ and $\dot\delta_0$ the opening and its rate. Throughout the analysis, both the initial state $\mathbf{x}_0$ and the time step $\Delta t$ are held fixed and chosen so that every studied trajectory crosses the interface $\delta=0$ in a single \acrshort{CD} step. Such a condition yields $\delta_1$ of opposite sign to $\delta_0$ and thus enforces a stiffness change. Consequently, the transition is parametrized only by the oriented stiffness ratio $\alpha=k_1/k_0$. Figure~\ref{fig:sdof_overshoot_schematic} illustrates the resulting traction/separation paths for various $\alpha$, highlighting the slope discontinuity at $\delta=0$.

\paragraph{Energy measure.}
Following \citep{acary2016energy}, the discrete‐energy invariant (or "algorithmic energy") for the explicit \acrshort{CD} scheme is
\begin{equation}
  \mathcal{H} \;=\; \mathcal{E}\;-\;\frac{\Delta t^2}{8}\,\mathbf{a}^\top\mathbf{M}\,\mathbf{a},
  \qquad
  \mathcal{E} \;=\; \mathcal{R} + \mathcal{K},
  \label{eq:algorithmic_energy_gen}
\end{equation}
where $\mathcal{R}$ and $\mathcal{K}$ are the reversible cohesive and kinetic energies, respectively.  
Under smooth evolution of a linear system, $\mathcal{H}$ remains exactly conserved (to machine precision), with the term in $\Delta t^2$, integration scheme dependent, eliminating the typical oscillations observed in the standard mechanical energy $\mathcal{E}$. This choice isolates the energy artifacts due solely to nonsmooth events. In the scalar SDOF case (mass $m$, stiffness $k$), Eq.~\ref{eq:algorithmic_energy_gen} reduces to
\begin{equation}
  \mathcal{H}
    = \frac12\,k\,\delta^2
    + \frac12\,m\,\dot\delta^2
    - \frac{\Delta t^2}{8}\,m\,\ddot\delta^2.
      \label{eq:algorithmic_energy}
\end{equation}

\noindent
Starting from $\mathbf{x}_0 = [\delta_0,\,\dot\delta_0]^\top$ with stiffness $k_0$, we compute $\mathcal{H}_0$. One explicit update yields the state $\mathbf{x}_1$ with $\delta_1$ of opposite sign to $\delta_0$, at which point the stiffness switches to $k_1 = \alpha\,k_0$. The algorithmic energy difference between $\mathcal{H}_1$ and $\mathcal{H}_0$ is given by
\begin{equation}
  \begin{aligned}
    \label{eq:energy_jump}
    \Delta\mathcal{H} &= \tfrac12(\alpha - 1)\,k_0\,\delta_0\delta_1 \\
        &= (\alpha - 1)\left[\tfrac12\,k_0\,\delta_0^2
        + \tfrac{\Delta t}{2}\,k_0\,\delta_0\,\dot\delta_0
        - \tfrac{\Delta t^2}{4m}\,k_0^2\,\delta_0^2\right],
  \end{aligned}
\end{equation}
using displacement update from Eq.~\ref{eq:central-difference-u} on $\delta_1$. This expression is a quadratic polynomial in $\Delta t$, proportional to $(\alpha-1)$. Since $\delta_0$ and $\delta_1$ have opposite signs, their product $\delta_0\delta_1$ is negative, implying
\[
  \mathrm{sign}(\Delta\mathcal{H}) \;=\; -\,\mathrm{sign}(\alpha - 1).
\]
Thus, if $\alpha>1$ ($k_1>k_0$), the algorithmic energy decreases (artificial dissipation), 
while if $\alpha<1$ ($k_1<k_0$), the algorithmic energy increases (artificial injection). 
Exact conservation is recovered for $\alpha=1$, as expected for a continuous system. 

\medskip\noindent

\begin{figure}[h!]
  \centering
  \includegraphics[scale=1]{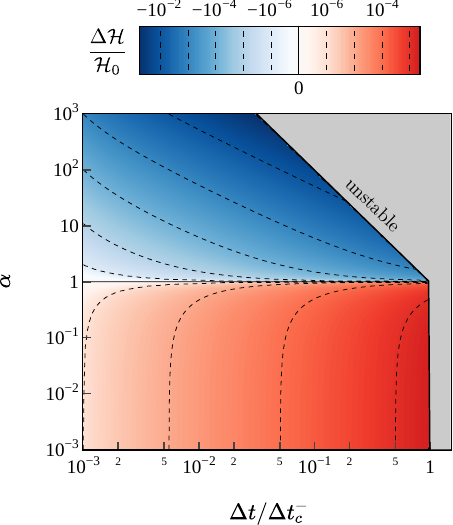}
  \caption{
    Heatmap of $\Delta\mathcal{H}/\mathcal{H}_0$ versus $\Delta t/\Delta t_c^-$ and $\alpha=k^+/k^-$. 
    Blue shading indicates energy dissipation ($\Delta\mathcal{H}<0$); red shading indicates energy injection ($\Delta\mathcal{H}>0$). 
    The dashed lines mark orders of magnitude of the energy jump. 
    The region in gray is marked as unstable, since the time step there exceeds the critical step of the stiffer branch.}
  \label{fig:sdof_overshoot_heatmap}
\end{figure}

\medskip\noindent
To visualize the behavior predicted by Eq.~\eqref{eq:energy_jump}, we fix $m=1$, $k^-=1$, and initialize the state $\mathbf{x}_0=[-10^{-3},\,1]^\top$, so that a single CD update drives the opening through zero and triggers the switch from $k_0=k^-$ to $k_1=k^+$. We sweep the time-step $\Delta t$ and the stiffness ratio $\alpha = k^+/k^-$, and, to present the results, we non-dimensionalize the time step by the contact critical time step
\begin{equation}
  \Delta t_c^- \;:=\; 2\sqrt{\frac{m}{k^-}} ,
  \label{eq:dtc_minus}
\end{equation}
defined with respect to the fixed penalty $k^-$. Figure~\ref{fig:sdof_overshoot_heatmap} reports the normalized algorithmic energy jump $\Delta\mathcal{H}/\mathcal{H}_0$ as a function of the normalized step $\Delta t/\Delta t_c^-$ and $\alpha$. On the map, blue regions (\(\Delta\mathcal{H}<0\)) correspond to \(\alpha>1\) (stiffer cohesion, net dissipation). In contrast, red regions (\(\Delta\mathcal{H}>0\)) correspond to \(\alpha<1\) (softer cohesion, net injection); the line at \(\alpha=1\) shows exact conservation (\(\Delta\mathcal{H}=0\)). In the gray region, the time step is larger than the system's critical time step; it is therefore labeled as unstable.

\medskip\noindent
The energy errors are computed with $\alpha=k^+/k^-$ for a contact$\to$cohesion transition and then a different $\alpha=k^-/k^+$ for a cohesion$\to$contact transition. Therefore, successive transitions alternately inject and dissipate energy. Whether these errors compensate or accumulate depends on the system parameters and the time step. Without external loading, the branch stiffnesses set the oscillation periods, so that the trajectory spends roughly half a period in each branch before switching. The alignment of these half-periods with the discrete time-step then controls the overshoot at each crossing and thus the net balance of injected and dissipated energy. We address this question in Section~\ref{sssec:phase_space_stability} by analyzing trajectories that repeatedly cross the cohesive–contact interface, combining phase-space diagnostics with energy-growth measures.

\subsubsection{Stability analysis under repeated transitions}
\label{sssec:phase_space_stability}
In Section \ref{sssec:one_step_energy_jump}, we showed that each crossing of a discontinuity in the traction/separation path introduces errors in the algorithmic energy. We now test whether repeated transitions can lead to unbounded energy growth, i.e., a global instability. The analysis uses the same single–degree–of–freedom (SDOF) surrogate (Figure \ref{fig:sdof_system}) advanced by the explicit \acrshort{CD} map. 

\medskip\noindent
For reference across this parametric study, we define the reference stiffness ratio of a system
\begin{equation}
  \alpha_{\mathrm{ref}} \;:=\; \frac{k^+}{k^-},
  \label{eq:alpha_ref_def}
\end{equation}
always anchored to the prescribed penalty $k^-$; hence $\alpha\in\{\alpha_{\mathrm{ref}},\,\alpha_{\mathrm{ref}}^{-1}\}$ depending on the transition direction. 

\medskip\noindent
Similarly to the previous section, we set the system with unit mass and unit contact penalty: $m=k^-=1$. The initial state is set to $\mathbf{x}_0=[\,0,\,1\,]^\top$ and we vary both the time-step $\Delta t$ and the cohesive stiffness $k^+$. We let the system evolve under free oscillation, allowing it to cross the discontinuity between cohesive and contact stiffness $200$ times. Equivalently, suppose we define a cycle as the interval between two consecutive upward zero-crossings of $\delta$, corresponding to the crossing from contact to cohesion. In that case, we let the system evolve for $n_\text{cycles}=100$ cycles. For each configuration $(k^+,\Delta t)$, we plot the trajectory $(\delta,\dot\delta)$ in a phase-space diagram against the iso–energy ellipse
\begin{equation}
  \mathcal{E} = \frac12\,k(\delta)\,\delta^2 + \frac12\,m\,\dot\delta^2,\qquad k(\delta)=\begin{cases}
      k^+, \quad \text{if } \delta > 0\\
      k^-, \quad \text{if } \delta < 0
  \end{cases},
  \label{eq:energy_sdof}
\end{equation}
splitted in two halves around $\delta=0$. Trajectories that remain close and oscillate around this contour $\mathcal{E}$ are classified as bounded; divergence indicates instability. In parallel, we track the normalized mechanical–energy variation $\Delta\mathcal{E}/\mathcal{E}_0$ as a function of cycles. 
\begin{figure*}[h!]
  \centering
  \includegraphics[width=\textwidth]{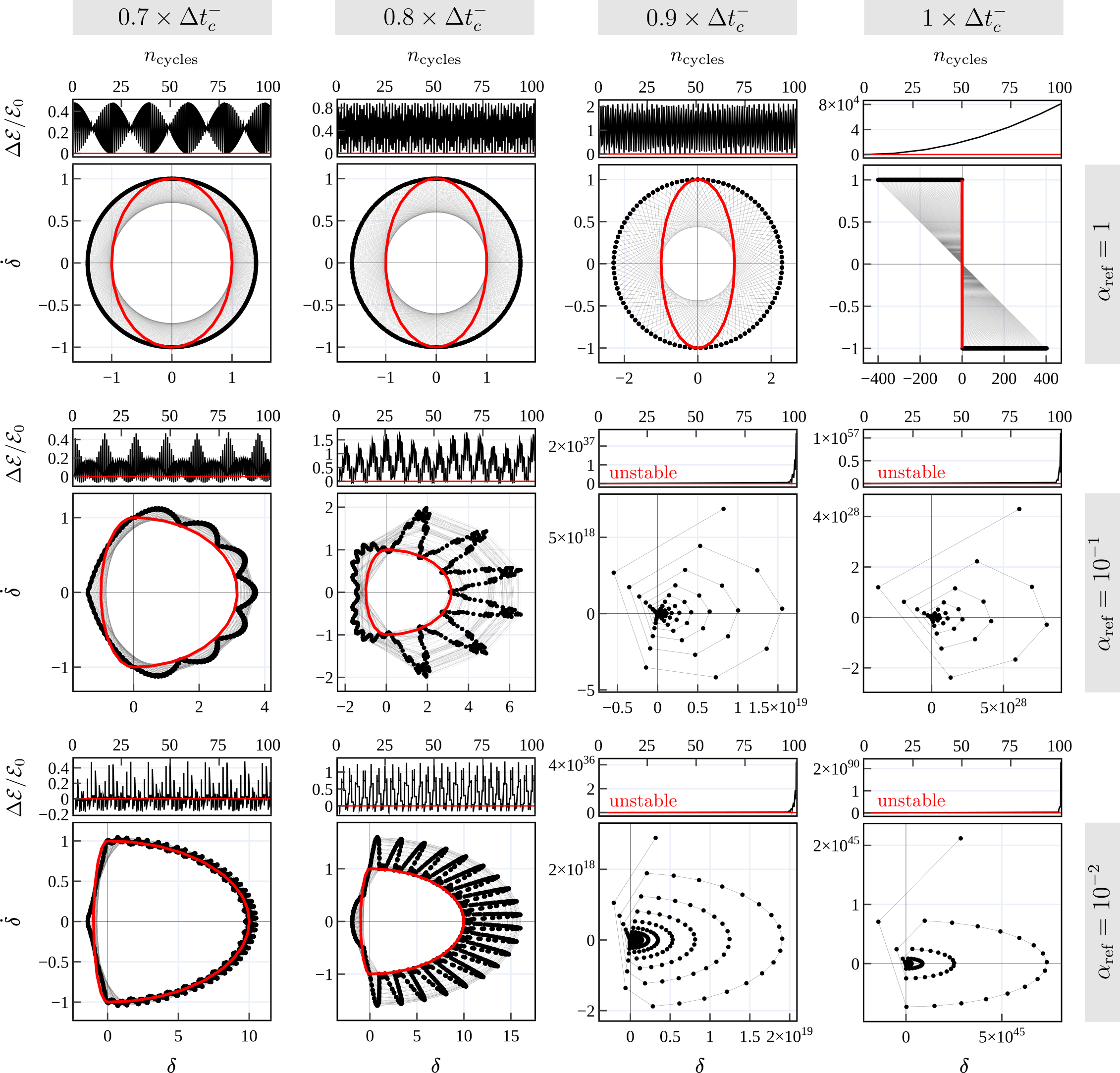}
  \caption{Phase-space trajectories for $\alpha_\text{ref}\in\{1,\,10^{-1},\,10^{-2}\}$ (rows) and $\Delta t/\Delta t_c^-\in\{0.7,\,0.8,\,0.9,\,1.0\}$ (columns). Black dots denote the discrete states computed by the integration scheme, connected by thin gray lines to indicate the trajectory. The red ellipse is the iso-energy curve $\tfrac12\,k\,\delta^2+\tfrac12\,m\,\dot\delta^2=\mathcal{E}$. This is also evident in the energy plot, which corresponds to the analytical phase-space trajectory/energy. Within each row, this curve is identical, with only the scale varying. For $\alpha_\text{ref}=1$ (no stiffness contrast), the classical stability bound $\Delta t\le\Delta t_c^-$ shows marginal stability, and decreasing $\Delta t/\Delta t_c^-$ contracts the loops toward the ellipse. For $\alpha_\text{ref}<1$ (stiffer contact than cohesion), outward drift appears at $\Delta t/\Delta t_c^-\gtrsim0.9$, whereas $\Delta t/\Delta t_c^-\le0.8$ produces bounded orbits.}
  \label{fig:sdof_phase-space}
\end{figure*}

\medskip\noindent
Results for $\alpha_\text{ref}\in\{1,\,10^{-1},\,10^{-2}\}$ and $\Delta t/\Delta t_c^-\in\{0.7,\,0.8,\,0.9,\,1.0\}$ (with $\Delta t_c^-$ from Eq.~\ref{eq:dtc_minus}) are presented in Figure~\ref{fig:sdof_phase-space}. Note that with $\alpha_\text{ref}<1$, $k^-$ is stiffer than $k^+$ and therefore $\Delta t_c^-$ should act as a true stability bound. For $k^+=k^-$ ($\alpha_{\mathrm{ref}}=1$), i.e., a single linear spring, the dynamics recover the expected behavior (top row of Fig.~\ref{fig:sdof_phase-space}): at $\Delta t=\Delta t_c^-$ the scheme is marginally stable, and any safety margin $\Delta t<\Delta t_c^-$ yields bounded, closed orbits. Also, consistent with what we expect, the \acrshort{CD} scheme conserves energy only in a weak sense: the mechanical energy oscillates about a constant mean, with oscillation amplitude decreasing as $\Delta t$ decreases. When $\alpha_{\mathrm{ref}}\neq 1$, a discontinuity is introduced in the traction/separation path, producing energy errors across each stiffness transition. For $\alpha_{\mathrm{ref}}<1$ (second and third rows), the standard stability condition is no longer sufficient: for $\Delta t \ge 0.9\,\Delta t_c^-$ the phase portraits develop outward spirals and the energy diverges, showing how many small energy errors can accumulate into a global instability. For $\Delta t<0.9\,\Delta t_c^-$, the trajectories and total energy remain bounded, though the phase-space loops exhibit complex orbits, indicating significant errors in displacement and velocity.

\medskip\noindent
Figure~\ref{fig:sdof_phase-space} thus shows that, for specific $(\alpha_{\mathrm{ref}},\Delta t)$ and initial state $\mathbf{x}_0$, repeated transition across a discontinuity can lead to unbounded energy growth even when the standard time step condition holds. It is expected: standard \acrshort{CD} stability bound is derived for linear time-invariant dynamics, yielding a fixed state-transition matrix $\mathbf{\Phi}$ (Section~\ref{ssec:time-integration}). In that case, stability is governed by the spectral radius $\rho(\mathbf{\Phi})=\rho(\mathbf{\Phi}^n)^{1/n}$, with $n$ combinations of the same $\mathbf{\Phi}$ matrix. Once different matrices update the state, as is the case for piecewise linear systems, stability is governed instead by the joint spectral radius \citep{rota1960note} of the set of update matrices. Because the switching sequence is trajectory-dependent and unknown a priori, a universal stability condition for such piecewise-linear dynamics is computationally intractable in general (see Appendix~\ref{app:sdof_switching}). It motivates a simpler finite–time stability survey over a wider parametric range.

\paragraph{Energy–peak growth–rate map.}
To map stability over \((\alpha_\text{ref},\,\Delta t/\Delta t_c^-)\) for $\mathbf{x}_0$ fixed, we quantify energy growth with an exponential envelope,
\[
\mathcal{E}(t) \sim Ce^{\lambda t}
\]
and estimate the energy-growth rate $\lambda \;[\text{s}^{-1}]$ from \emph{per-cycle} peaks to isolate the energy envelope. Cycles are delimited by upward zero-crossings of the opening $\delta$: for each time interval \([\tau_j,\tau_{j+1})\) delimiting cycle $j$,
\[
\mathcal{E}_{\rm peak}(j):=\max_{\tau_j<t<\tau_{j+1}}\mathcal{E}(t).
\]
Assuming \(\log \mathcal{E}_{\rm peak}(t)\) is approximately linear in \(t\), we fit its slope $\lambda$ with the Theil–Sen estimator \citep{theil1950rank,sen1968estimates}. Analytically, stability would be set by $\lambda=0$, when energy is conserved. However, to account for finite-time uncertainty, we define stability by comparing $\lambda$ to a tolerance $\epsilon_\lambda$: stable if $\lambda \le \epsilon_\lambda$, unstable otherwise. Calibrating $\epsilon_\lambda$ on the linear reference $\alpha_\text{ref}=1$ gives a numerical floor $\lambda\approx 10^{-6} \text{ s}^{-1}$ for $\Delta t <\Delta t_c^-$ and genuine growth $\lambda > 10^{-2}\text{ s}^{-1}$ for $\Delta t \ge \Delta t_c^-$; we therefore adopt $\epsilon_\lambda=2\times 10^{-3}\text{ s}^{-1}$ as a conservative cutoff.

\begin{figure}[h!]
  \centering
  \includegraphics[scale=1]{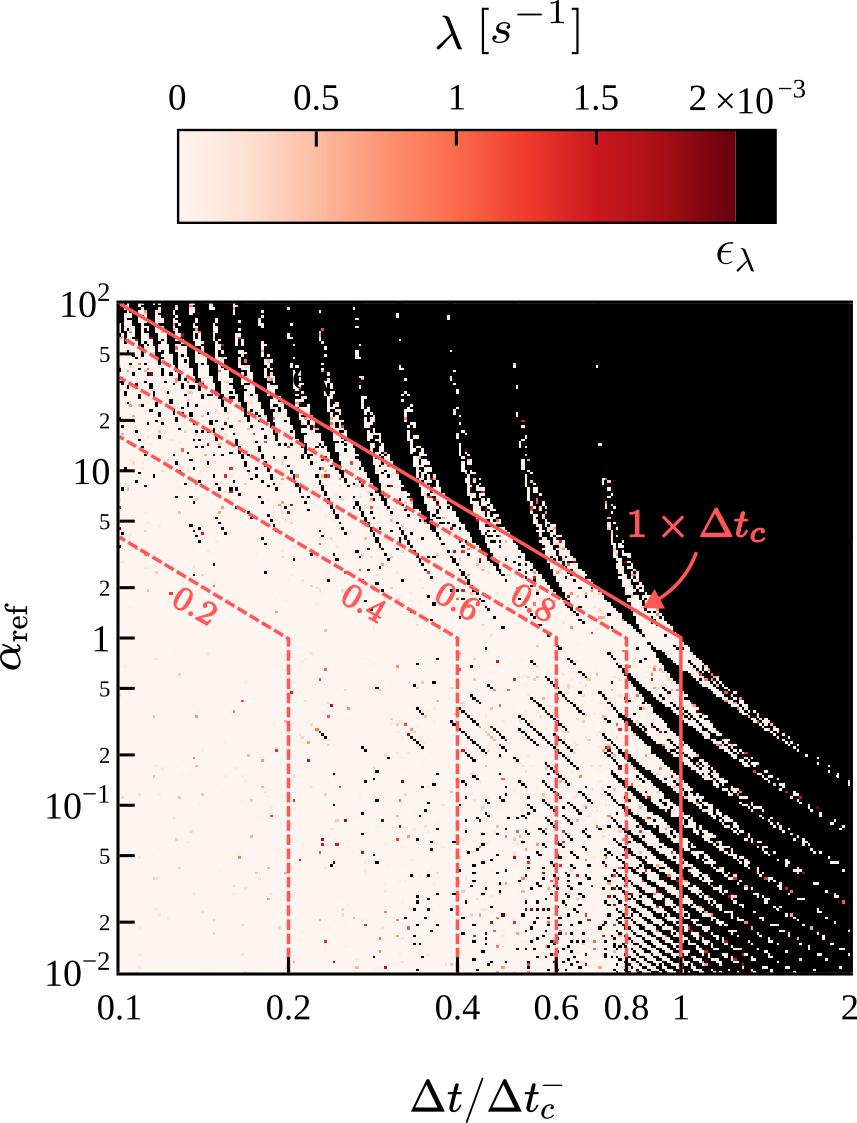}
  \caption{
    Map of the energy growth rate $\lambda$ over a $256\times256$ grid in the $(\alpha_\text{ref},\Delta t/\Delta t_c^-)$ plane. Cells are colored by $\lambda$ on a linear color scale; values $\lambda \ge \epsilon_\lambda$ (declared unstable) are shown in black. The bold red line mark the classical stability limit $\Delta t_c$ expressed in the chosen normalization with $\Delta t_c=\Delta t_c^-$ when $\alpha_\text{ref}<1$ and $\Delta t_c=\Delta t_c^+$ when $\alpha_\text{ref}>1$. Dashed red line indicate fractions of this critical time step, equivalent to adding a safety factor. Note sparse stability windows beyond $\Delta t_c$ and pronounced instability below it.
  }
  \label{fig:sdof_growth_map}
\end{figure}

\medskip\noindent
Figure~\ref{fig:sdof_growth_map} shows a dense scan in the \((\alpha_\text{ref},\,\Delta t/\Delta t_c^-)\) plane on a \(256\times256\) grid, with \(\alpha_\text{ref}\in[10^{-2},\,10^2]\) and \(\Delta t/\Delta t_c^-\in[0.1,\,2]\). Each cell represents a 200-cycle simulation, colored by the estimated energy growth rate $\lambda$. The bold red curve marks the branchwise stability limit $\Delta t_c\in\{\Delta t_c^-, \Delta t_c^+\}$ and dashed lines show fractions of this critical time step. Only at $\alpha_\text{ref}$ does this limit coincide with an actual stability boundary; for $\alpha_\text{ref}\neq 1$, stiffness switching makes the classical stability limit non-predictive -- unstable regions appear even for $\Delta t<\Delta t_c$, while stable regions persist beyond it. The map exhibits ripple-like bands of alternating stability, highlighting a strongly non-monotonic dependence on the initial state $\mathbf{x}_0$, the stiffness contrast $\alpha_\text{ref}$, and the time-step.

\paragraph{Practical implications}
In full-scale fragmentation simulations, interfaces frequently transition between cohesive and contact states, exhibiting significant stiffness contrasts. Figure~\ref{fig:sdof_growth_map} implies that:
\begin{itemize}
  \item For \(\alpha_\text{ref}\approx1\) or infrequent switching, the classical limit on the time-step $\Delta t_c$ can be considered as acceptable.
  \item For frequent switching at \(\alpha_\text{ref}\neq 1\), time-steps must be reduced below \(\Delta t_c\) by a safety factor drawn from the $\lambda$ map to ensure $\lambda<\epsilon_\lambda$; in the present SDOF setup, a conservative choice is \(\Delta t \lesssim 0.2\,\Delta t_c\).
\end{itemize}
This \acrshort{SDOF} study isolates and quantifies the cohesive–contact source of instability, but it does not yield a universal step-selection rule. The stability boundaries are specific to the surrogate system, the initial state $\mathbf{x}_0$, and the explored parameter ranges; multi-degree-of-freedom systems or different initial conditions may shift these boundaries.

\medskip\noindent
In the following subsection, we turn to another potential source of instability: \emph{softening} events introduce a second discontinuity in the traction-separation law, from which we also expect energy errors. We assess its impact on the stability and energy balance of the explicit \acrshort{CD} integrator in the next section.

\subsection{Softening}
\label{ssec:softening}

\noindent
In the present cohesive–zone formulation, \emph{softening} refers to the progressive reduction of the cohesive stiffness \(k^+\) once the opening \(\delta\) exceeds its current peak value \(\delta_{\max}\). As \(\delta\) grows from \(\delta_0\le \delta_{\max}\) to \(\delta_1>\delta_{\max}\), the stiffness degrades from its initial value \(k_0\) down to \(k_1 < k_0\), as shown in Figure~\ref{fig:instabilities_tsl}c.  This damage‐dependent law is designed to capture the material's loss of load‐carrying capacity during fracture.

\medskip\noindent
At the cohesive-contact transition, the energy balance should ideally satisfy $\Delta\mathcal{H}=0$; any deviation is considered as a numerical error. During softening, however, dissipation is expected, so the full balance becomes $\Delta\mathcal{H}+\Delta\mathcal{G}=0$, where $\Delta\mathcal{G}$ denotes the fracture dissipation term. Since softening introduces another discontinuity in the \acrshort{TSL}—previously shown to generate numerical error—it is necessary to verify whether this balance holds. By analogy the \acrshort{SDOF} study, $\Delta\mathcal{H}$ is evaluated using Equation~\ref{eq:energy_jump}, with $\alpha=k_1/k_0<1$. This jump is strictly negative since $k_0\delta_0\delta_1>0$ and $\alpha-1<0$. \(\Delta \mathcal{G}\) is defined as the difference between the work done by the cohesive tractions $W_{0\to1}^\text{coh}$ and the change in stored cohesive energy $\Delta \mathcal{R}$:
\begin{equation}
\Delta\mathcal{G} = W^{\mathrm{coh}}_{0\to1} - \Delta\mathcal{R}.
\label{eq:dissipation_increment_app}
\end{equation}
where,
\begin{equation}
  \Delta\mathcal{R} = \mathcal{R}_1 - \mathcal{R}_0
    = \frac12\,k_1\,\delta_1^2 - \frac12\,k_0\,\delta_0^2,
\label{eq:elastic_energy_change_app}
\end{equation}
and,
\begin{equation}
  \begin{aligned}
  W^{\mathrm{coh}}_{0\to1} &= (\delta_1 - \delta_0)f^{\mathrm{coh}}_{1/2}\\
  &= \frac{1}{2}(\delta_1 - \delta_0)(k_0\delta_0 + k_1\delta_1)\\
  \end{aligned}
\label{eq:cohesive_work_app}
\end{equation}
Subtracting the two terms results in the dissipation increment being expressed as
\begin{equation}
\begin{aligned}
  \Delta\mathcal{G} 
        &= \frac{1}{2}(1-\alpha)k_0\delta_0\delta_1
\end{aligned}
\label{eq:dissipation_increment_final_app}
\end{equation}
Because \(\alpha<1\), \(\Delta\mathcal{G}>0\), satisfying the condition of positive-only dissipation.  Comparing Eqs. \ref{eq:energy_jump} and \ref{eq:dissipation_increment_final_app} confirms
\begin{equation}
\Delta\mathcal{H} + \Delta\mathcal{G} = 0,
\label{eq:energy_balance_final_app}
\end{equation}
While softening events are characterized by a nonsmooth path in the traction–separation law, they neither artificially inject nor dissipate energy, unlike the cohesive/contact transition. The energy lost in the algorithmic energy \(\Delta\mathcal{H}<0\) is exactly offset by the fracture dissipation \(\Delta\mathcal{G}>0\), leading to a net zero numerical energy error. The softening events are therefore numerically conservative and \emph{benign} from a stability standpoint.

\section{Discussion and Outlook}
\label{sec:discussion}

\noindent
Section~\ref{sec:instability-mechanisms} examined three sources of instability that could explain the energy drift and the late-time growth of fragment count reported in Section~\ref{ssec:observed_instability}. We find that at softening events, the loss of algorithmic energy is exactly balanced by an increase in fracture dissipation and therefore does not, by itself, provoke instability. By contrast, the high initial stiffness of newly inserted cohesive elements and the abrupt stiffness switch at contact pose a more substantial risk.

\medskip\noindent
In our test run, the maximal observed cohesive stiffness did not exceed the admissible limit set by the time step and was therefore not the source of instability. However, under lower or spatially non-uniform strain rates, it can become a decisive factor. A practical delimitation of the unstable region is provided by a minimal admissible damage threshold \(\tilde d\) (equivalently, a maximal stiffness \(\tilde k^+\)), below which the explicit time-step would violate stability. This region can be eliminated by locally modifying the \acrshort{TSL}: by capping the cohesive stiffness to a fixed bound or by enforcing a constant-traction unload/reload path as in \citep{collins2022formulation} and \citep{zhou2005cohesive}. Numerically, such strategies prevent low-damage interfaces from imposing vanishingly small time-steps. However, when combined with a penalty-based contact law, these modifications prevent the path followed by the \acrshort{TSL} from reaching the origin $(0,0)$, where the contact law begins. As a result, instead of a slope discontinuity at the transition, one introduces an actual traction jump at $\delta=0$, thus losing $\mathcal{C}^0$ continuity. It can therefore be expected to exacerbate instabilities. Any such remedy should consequently be paired with a more robust treatment of contact.

\paragraph{Adaptive Contact Penalty.} The most critical source of instability is the cohesive–contact transition. As established in Section~\ref{ssec:cohesive-contact}, repeated crossings of \(\delta=0\) generate alternating energy injections and dissipation that can accumulate and lead to an energy drift. To isolate this effect on the global instability identified in Section~\ref{ssec:cohesive-contact}, we replace the fixed contact penalty \(k^-\) with a damage‐dependent variant following the cohesive stiffness, i.e.\  
\[
  k^- \;=\; k^+(d)\quad\text{(cf.\ Eq.~\ref{eq:k_plus_damage})},
\]  
thereby restoring continuity at \(\delta=0\) and suppressing any energy errors. We run the same simulation as in Section \ref{ssec:problem-setup} with this modification of contact to confirm that cohesive-contact transitions are the leading instability mechanism, also in this multi-dimensional problem. All other modeling parameters --triangular elements of size \(h=3.7\times10^{-5}\,\mathrm{m}\), time-step \(\Delta t=0.4\,\Delta t_{c,G}\) (with \(\Delta t_{c,G}\) still computed using the originally fixed \(k^-\)), and aluminum oxide material properties--are identical, ensuring a one‐to‐one comparison.

\begin{figure*}[h!]
    \centering
    \includegraphics[scale=1]{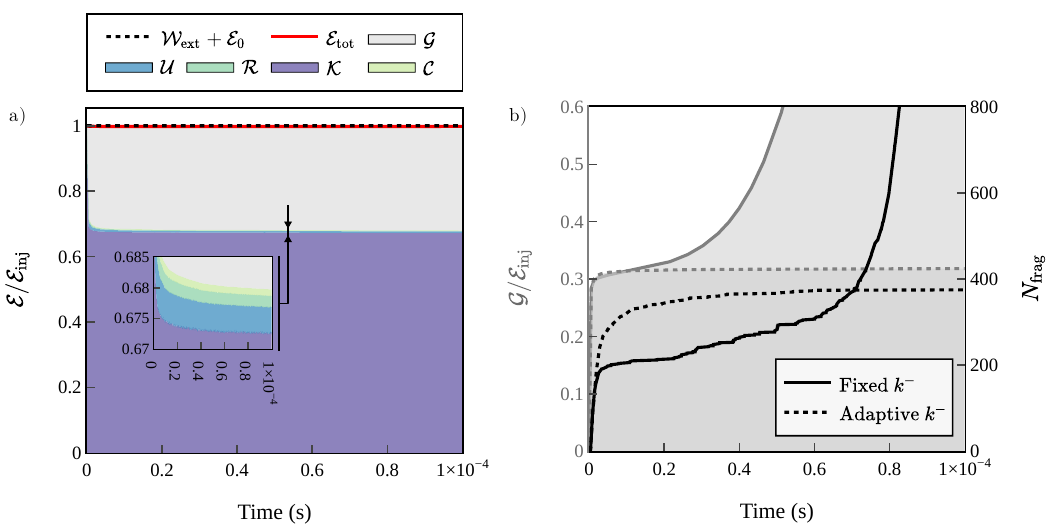}
    \caption{
        \textbf{(a)} Energy partitioning with the condition \(k^- = k^+(d)\), which removes the nonsmoothness associated with cohesive/contact transitions. This adjustment leads to stable energy evolution, with internal plus dissipated energy precisely tracking the external work.
        \textbf{(b)} Comparison between the reference case (solid lines) and the adaptive-\(k^-\) approach (dashed lines). The black curves represent the number of fragments, indicating that the adaptive-\(k^-\) strategy results in a stable plateau. The filled curves depict the evolution of fracture energy \(\mathcal{G}\). Although the adaptive-\(k^-\) case results in a slightly higher final fragment count, both configurations evolve similarly in terms of \(\mathcal{G}\) during the main fragmentation event.
    }
    \label{fig:frag2D_adaptivecontact_2}
\end{figure*}

\medskip\noindent
Figure~\ref{fig:frag2D_adaptivecontact_2} presents the results. In panel~(a), the sum of internal energy components (recoverable + kinetic) with dissipation precisely tracks the injected energy with no discernible drift, confirming that the global energy balance is restored once the stiffness discontinuity is removed. Panel~(b) overlays the fragment count \(N_{\rm frag}\) and fracture energy \(\mathcal{G}\) for both the reference (solid lines) and adaptive‐penalty (dashed lines) cases.  The adaptive contact penalty simulation reaches and maintains a stable fragment‐count plateau, despite a higher final \(N_{\rm frag}\), and follows a similar \(\mathcal{G}\) evolution as the reference case through the main fragmentation event.

\medskip\noindent
While this approach restores energy balance, non-physical interpenetration becomes more pronounced at higher cohesive damage levels, since the contact stiffness decays with increasing damage. Consequently, fragment statistics, although numerically stable, cannot be regarded as physically accurate. While this approach serves as a valuable diagnostic tool to isolate the effect of the cohesive-contact switch, it cannot be viewed as a viable long-term remedy for practical fragmentation simulations.

\medskip\noindent
Penalty methods face an inherent trade-off between preventing interpenetration, maintaining feasible time steps, and controlling energy behavior, which cannot be resolved in a problem-independent way. Heuristic fixes (e.g., local time-step reduction, viscous or bi-penalty stabilization, mass redistribution) can mitigate symptoms. Nonsmooth time integrators offer a more consistent alternative, enforcing contact via hard velocity-level constraints and impulses \citep{chen2012newmark,chen2013nonsmooth}. Such schemes align with the intrinsic nonsmooth nature of contact mechanics and ensure clear energy properties across impacts \citep{acary2016energy}. Although each step may be costlier than a purely explicit penalty update, the improved stability could allow time steps that stiff penalties would otherwise forbid; stability limits then stem from bulk and cohesion, rather than from contact. This approach also enables cohesive stiffness capping, or constant-traction unload/reload paths, without introducing a discontinuity at $\delta=0$, thereby addressing the high initial stiffness source as well.

\section{Conclusion}
\label{sec:conclusion}

\noindent
We examined the stability of dynamic fragmentation simulations that couple an extrinsic \acrshort{CZM} with penalty-based contact under explicit \acrlong{CD} integration. By isolating the sources of instability in simplified settings, we established a mechanism-aware view of when and why energy grows and artificial fragmentation occurs. Three potential sources were identified and quantified: 
\begin{enumerate}
  \item \textbf{Diverging initial stiffness.} As the damage $d$ tends to zero, the cohesive stiffness \(k^+\) can become arbitrarily large, so no single global \(\Delta t\) uniformly resolves the resulting transients. A spectral radius analysis yields a damage threshold \(\tilde d(\Delta t)\) below which instability arises.
  \item \textbf{Cohesive–contact switching.} Penalty-based contact enforces non-penetration weakly via a fixed stiffness \(k^-\). The abrupt transition between the cohesive stiffness \(k^+\) and a fixed contact penalty \(k^-\) introduces a discontinuity in the force–displacement response at \(\delta=0\) that generates energy errors; under repeated transitions, these errors can accumulate into energy drift and fragment–count inflation, even under standard stability conditions.
  \item \textbf{Gradual softening.} Softening events introduce a non–smooth stress path similar to the cohesive–contact transition. However, the resulting energy loss is balanced exactly by the dissipation increment; by itself, it does not trigger instability.
\end{enumerate}

\medskip\noindent
Among these, the cohesive–contact switch is the dominant source of instability, exposing an inherent limitation of penalty-based contact in fragmentation simulation. A diagnostic variant with an adaptation of the contact penalty \(k^-(d)=k^+(d)\) removes the discontinuity and restores energy balance, but at the cost of damage–dependent interpenetration; it is therefore not a viable remedy. The broader implication is clear: penalty-based contact forces a compromise between preventing interpenetration (large \(k^-\)) and maintaining acceptable time-steps and energy behavior; this compromise is problem–dependent and cannot be tuned away, making this approach unsuitable for robust, long-term simulation of fragmentation with \acrshort{CZM} in explicit.

\medskip\noindent
We therefore advocate for nonsmooth time-stepping schemes that enforce unilateral contact at the velocity level, eliminating the artificial stiffness jump while shifting the time step restriction back to bulk and cohesive dynamics. Future work will integrate and benchmark such a scheme against penalty-based contact, assessing stability, and validating its predictive fidelity and scalability on realistic fragmentation scenarios.

\subsection*{Acknowledgements}
This research was funded by the Swiss National Science Foundation (SNSF) [Grant N° 212935].

\subsection*{Declaration of generative AI and AI-assisted technologies in the manuscript preparation process}
During the preparation of this work, the authors used ChatGPT 5 to improve the grammar and clarity of the manuscript. After using this tool/service, the authors reviewed and edited the content as needed and take full responsibility for the content of the published article.

\appendix
\onecolumn    
\appendix
\section{Derivation of the Central‐Difference State‐Transition Matrix and Stability Bound}
\label{app:cd_stability}

In this appendix, we derive the state‐transition matrix for the explicit central‐difference (CD) scheme and recall how the conditional stability bound \(\Delta t_c = 2/\omega_{\max}\) arises \citep{belytschko2014nonlinear}. 

\medskip\noindent
We collect displacements and velocities into the state vector
\[
  \mathbf{x} = \begin{bmatrix}\mathbf{u} \\ \mathbf{v}\end{bmatrix},
  \quad
  \mathbf{u}, \mathbf{v}\in\mathbb{R}^n.
\]
The semi‐discrete equations of motion for mass matrix \(\mathbf{M}\), stiffness matrix \(\mathbf{K}\), and external force \(\mathbf{f}^{\rm ext}\) is defined in Equation~\ref{eq:semi-discrete-motion}. Applying the \acrshort{CD} update (Newmark–\(\beta\) with \(\beta=0,\gamma=\tfrac12\), see Eq.~\ref{eq:central-difference-u}-\ref{eq:central-difference-a}) to the undamped, unforced case, between steps $n$ and $n+1$ yields
\begin{align}
  \mathbf{u}_{n+1} &= \mathbf{u}_n + \Delta t\,\mathbf{v}_n + \tfrac12\Delta t^2\,\mathbf{a}_n,\label{eq:cd_u}\\
  \mathbf{v}_{n+1} &= \mathbf{v}_n + \tfrac12\Delta t\,(\mathbf{a}_n + \mathbf{a}_{n+1}),\label{eq:cd_v}
\end{align}
with
\[
  \mathbf{a}_n = \mathbf{M}^{-1}\bigl(-\mathbf{K}\,\mathbf{u}_n\bigr).
\]
Substituting \(\mathbf{a}_n\) and \(\mathbf{a}_{n+1}\) into \eqref{eq:cd_u}–\eqref{eq:cd_v} expresses \(\mathbf{x}_{n+1}\) linearly in terms of \(\mathbf{x}_n\). After algebraic manipulation, one obtains
\begin{equation}
  \mathbf{x}_{n+1}
  = \underbrace{
    \begin{bmatrix}
      \mathbf{I} - \tfrac12\Delta t^2\,\mathbf{M}^{-1}\mathbf{K}
      & \Delta t\,\mathbf{I} \\
      -\Delta t\,\mathbf{M}^{-1}\mathbf{K}
      + \tfrac14\Delta t^3\,\mathbf{M}^{-1}\mathbf{K}\,\mathbf{M}^{-1}\mathbf{K}
      & \mathbf{I} - \tfrac12\Delta t^2\,\mathbf{M}^{-1}\mathbf{K}
    \end{bmatrix}
  }_{\mathbf{\Phi}}
  \mathbf{x}_n.
  \label{eq:phi}
  \end{equation}
Here \(\mathbf{\Phi}\in\mathbb{R}^{2n\times2n}\) is the state‐transition matrix for one time-step. The CD scheme is stable if and only if all eigenvalues \(\lambda_i\) of \(\mathbf{\Phi}\) satisfy \(\lvert\lambda_i\rvert\le1\), i.e.\ the spectral radius
\[
  \rho(\mathbf{\Phi}) = \max_i\lvert\lambda_i\rvert \le 1.
\]
A direct spectral analysis shows that this condition reduces to
\[
  \Delta t \;\le\; \frac{2}{\omega_{\max}},
  \quad
  \omega_{\max} = \sqrt{\lambda_{\max}\!\bigl(\mathbf{M}^{-1}\mathbf{K}\bigr)}.
\]
Here \(\lambda_{\max}(\mathbf{M}^{-1}\mathbf{K})\) is the largest eigenvalue of the generalized eigenproblem \(\mathbf{K}\,\mathbf{u} = \omega^2\mathbf{M}\,\mathbf{u}\).

\section{SDOF Extension for Piecewise‐Linear Stiffness}
\label{app:sdof_switching}

Building on Appendix~\ref{app:cd_stability}, we now specialize the general CD state‐transition matrix \(\mathbf{\Phi}\) to the single‐degree‐of‐freedom case and derive the switched‐stiffness maps.

\medskip\noindent
From Eq.~\ref{eq:phi}, setting \(\mathbf{M}=m\) and \(\mathbf{K}=k\) yields the 2×2 map
\begin{equation}
  \mathbf{\Phi}(\Delta t,k)
  = 
  \begin{bmatrix}
    1 - \frac12\Delta t^2\,\frac{k}{m} & \Delta t \\
    -\Delta t\,\frac{k}{m} + \frac14\Delta t^3\,\frac{k^2}{m^2}
      & 1 - \frac12\Delta t^2\,\frac{k}{m}
  \end{bmatrix}.
\end{equation}
Let the stiffness alternate between the cohesive value \(k^+\) and the contact value \(k^-\). We define the constant‐stiffness maps
\[
  \mathbf{\Phi}^+ = \mathbf{\Phi}(\Delta t,\,k^+),
  \qquad
  \mathbf{\Phi}^- = \mathbf{\Phi}(\Delta t,\,k^-),
\]
and the two mixed‐stiffness maps by using \(k^+\) in the \(n\)-step acceleration and \(k^-\) in the \((n+1)\)-step acceleration (or vice versa):
\begin{equation}
  \mathbf{\Phi}^{+-}
  = 
  \begin{bmatrix}
    1 - \frac12\Delta t^2\,\frac{k^+}{m} & \Delta t \\
    -\frac{\Delta t}{2m}\,(k^+ + k^-)
      + \frac{\Delta t^3}{4m^2}\,(k^+ k^-)
      & 1 - \frac12\Delta t^2\,\frac{k^-}{m}
  \end{bmatrix},
  \label{eq:Phi_plusminus}
\end{equation}
\begin{equation}
  \mathbf{\Phi}^{-+}
  = 
  \begin{bmatrix}
    1 - \frac12\Delta t^2\,\frac{k^-}{m} & \Delta t \\
    -\frac{\Delta t}{2m}\,(k^- + k^+)
      + \frac{\Delta t^3}{4m^2}\,(k^- k^+)
      & 1 - \frac12\Delta t^2\,\frac{k^+}{m}
  \end{bmatrix}.
  \label{eq:Phi_minusplus}
\end{equation}

\medskip\noindent 
The per‐step update in the SDOF switched system is then
\[
  \mathbf{x}_{n+1}
  = 
  \begin{cases}
    \mathbf{\Phi}^+\,\mathbf{x}_n,  & \text{cohesion}\to\text{cohesion},\\
    \mathbf{\Phi}^-\,\mathbf{x}_n,  & \text{contact}\to\text{contact},\\
    \mathbf{\Phi}^{+-}\,\mathbf{x}_n, & \text{cohesion}\to\text{contact},\\
    \mathbf{\Phi}^{-+}\,\mathbf{x}_n, & \text{contact}\to\text{cohesion}.
  \end{cases}
\]
The overall map is an ordered product of these four matrices. The joint spectral radius determines stability \citep{rota1960note}. For the set of matrices $\mathcal{M}=\{\mathbf{\Phi}^+,\mathbf{\Phi}^-,\mathbf{\Phi}^{+-},\mathbf{\Phi}^{-+}\}$, it is defined by:
\[
    \rho(\mathcal{M}) =  \lim_{k\to\infty}
    \max\Bigl\{\|\mathbf{A}_{i_1}\cdots\mathbf{A}_{i_k}\|^{1/k}:\mathbf{A}_i\in\mathcal{M}\Bigr\}
  <1.
\]
Without prior knowledge of the combination of matrices, finding such a condition on stability is numerically intractable (NP-hard).       

\twocolumn
\bibliography{SWPaper}

\end{document}